%% file: main.tex
\RequirePackage{lineno}

\documentclass[twocolumn,showpacs,aps,prd,superscriptaddress]{revtex4}

\usepackage{graphicx}
\usepackage{dcolumn}
\usepackage{amsmath}
\usepackage{epsfig}

\input babarsym

\newcommand{\SLACPubNumber} {15687}

\begin{document}

\begin{flushleft}
\babar-PUB-13/013\\
SLAC-PUB-\SLACPubNumber\\[10mm]
\end{flushleft}

\title{
{\large \bf
Search for a light Higgs boson decaying to two gluons or $\mathbf{\ssbar}$ in the radiative decays of $\mathbf{\OneS}$ 
} 
}

\input authors_jun2013.tex

\begin{abstract}
\input abstract.tex
\end{abstract}


\maketitle

The next-to-minimal supersymmetric standard model (NMSSM), one of several extensions to the Standard Model \cite{ref:NMSSMreview}, predicts that there are two charged, three neutral $CP$-even, and two neutral $CP$-odd Higgs bosons. One of the $CP$-odd Higgs bosons, $A^0$, can be lighter than two bottom quarks \cite{ref:radups}.  If so, a $CP$-odd Higgs boson that couples to bottom quarks could be produced in the radiative decays of an $\Upsilon$ meson.

The $A^0$ is a superposition of a singlet and a non-singlet state. The branching fraction $\mathcal{B} (\Upsilon \to \gamma A^0)$ depends on the NMSSM parameter $\cos\theta_A$, which is the non-singlet fraction. The final state to which the $A^0$ decays depends on various parameters such as $\tan\beta$ and the $A^0$ mass \cite{ref:update}. \babar\ has searched for an $A^0$ decaying into \mumu~\cite{ref:mumu,ref:mumu2}, \tautau \cite{ref:tautau,ref:tautau2}, invisible states \cite{ref:invisb}, and hadronic final states \cite{ref:higgs2hadrons}, and has not seen a significant signal. The CMS collaboration has also not observed a significant signal in the search for $A^0$ decaying into \mumu \cite{ref:CMS}. In this article, we report on the first search for the decay $\OneS \to \gamma A^0 , A^0 \to gg$ or \ssbar. We search for the $A^0$ in the mass range $0.5 < m_{A^{0}} < 9.0\gevcc$. By tagging the dipion in the $\TwoS \to \pip\pim\OneS$ transition, this analysis greatly reduces $\epem\to\qqbar$ background, where $q$ is a $u,d,$ or $s$ quark, which is a dominant background contribution in \babar 's previous $A^0\to$ hadrons analysis \cite{ref:higgs2hadrons}. Although this analysis has been motivated by NMSSM, these results are generally applicable to any $CP$-odd hadronic resonances produced in the radiative decays of \OneS because we search for the $A^0$ excluding two-body final states. For an $A^0$ mass less than  $2m_\tau$, the $A^0$ is predicted to decay predominantly into two gluons if $\tan\beta$ is of order 1, and into \ssbar if $\tan\beta$ is of order 10.

This article uses data recorded with the \babar\ detector at the \pep2\ asymmetric-energy \epem\ collider at the SLAC National Accelerator Laboratory.  The \babar\ detector is described in detail elsewhere \cite{ref:detector,ref:detector2}. For this analysis, we use 13.6\invfb of data \cite{Lees:2013rw} taken at the \TwoS resonance (``on-resonance"). An estimated number of $(98.3 \pm 0.9)\times 10^6$ \TwoS mesons were produced. The branching fraction $\mathcal{B}(\TwoS \to \pip\pim\OneS)$ is ($17.92 \pm 0.26$)\% \cite{ref:pdg}. Therefore, $(17.6 \pm 0.3)\times 10^6$ \OneS mesons were produced via the dipion transition. We also use 1.4\invfb of data \cite{Lees:2013rw} taken 30\mev below the \TwoS resonance (``off-resonance") as a background sample.

Simulated signal events with various $A^0$ masses ranging from 0.5 to 9.0\gevcc are used in this analysis. The \evtgen event generator \cite{ref:evtgen} is used to simulate particle decays. The $A^0$ is simulated as a spin-0 particle decaying to either $gg$ or \ssbar. Since the width of the $A^0$ is expected to be much less than the invariant-mass resolution of $\approx$100\mevcc, we simulate the $A^0$ with a 1\mevcc decay width. \mbox{\tt Jetset} \cite{ref:jetset} is used to hadronize partons, and \textsc{Geant4} \cite{ref:geant4} is used to simulate the detector response.

\begin{figure}
\includegraphics[width=\columnwidth]{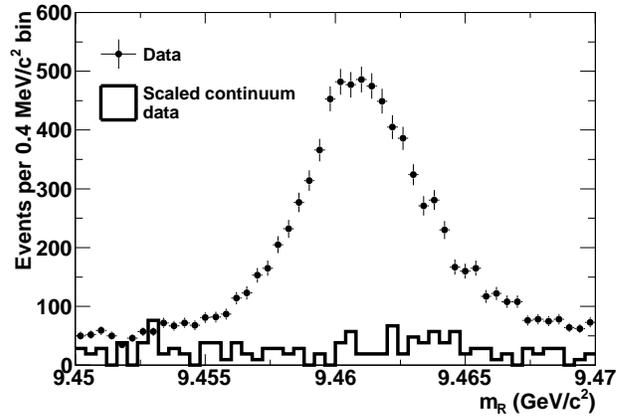}
\caption{
Distribution of the recoil mass against the dipion system in on-resonance data (points with error bars) after applying all selection criteria. The histogram is the continuum background recoil mass distribution from off-resonance data normalized to the on-resonance integrated luminosity.
}
\label{fig:dp}
\end{figure}

We select events with two charged tracks as the dipion system candidate, a radiative photon, and a hadronic system, as described later in this article. We select $\TwoS \to \pip\pim\OneS$ candidates based on the invariant mass $m_{R}$ of the system recoiling against the dipion system:
\begin{equation}
m_{R}^2 = M_{\TwoS}^2 + m_{\pi\pi}^2 - 2M_{\TwoS}E_{\pi\pi}^{CM},
\end{equation}

\noindent
where $M_{\TwoS}$ is the world average \TwoS mass \cite{ref:pdg}, $m_{\pi\pi}$ is the measured dipion invariant mass, and $E_{\pi\pi}^{CM}$ is the dipion energy in the \epem center-of-mass (CM) frame. The recoil mass distribution from an $\TwoS \to \pip\pim\OneS$ transition has a peak near the \OneS mass of $9.46030\pm0.00026\gevcc$ \cite{ref:pdg}. The background recoil mass distribution is uniform. We select events with a recoil mass in the range 9.45 to 9.47\gevcc . We further suppress the background with a multi-layer perceptron (MLP) neural network \cite{ref:TMVA}. Using simulated $\TwoS \to \pip\pim\OneS$, $\OneS \to \gamma A^0$ decays of various $A^0$ masses, \TwoS decays without dipions in the final state, and $\epem \to \qqbar$ events, we train an MLP using nine dipion kinematic variables \cite{ref:invisb}. The variables are: opening angle between the pions; absolute value of the cosine of the angle formed between the \pim and the direction of the \TwoS in the dipion frame; dipion momentum perpendicular to the beam axis; dipion invariant mass; distance from the beam spot; the larger momentum of the two pions; cosine of the dipion polar angle; $\chi^{2}$ probability of the fit of the two pion tracks to a common vertex; and cosine of the polar angle of the more energetic pion. These quantities are calculated in the \epem CM frame unless otherwise specified. Applying all other selection criteria, 99\% of the remaining signal events and 80\% of continuum events pass our MLP selection. The distribution of the recoil mass against the dipion system in data after applying all selection criteria is shown in Fig. \ref{fig:dp}.

\begin{table}[h]
  \begin{center}
    \caption[Decay modes for candidate $A^0\to gg$ and \ssbar decays, sorted by the total mass of the decay products.]{Decay modes for candidate $A^0\to gg$ and \ssbar decays, sorted by the total mass of the decay products.}
    \begin{tabular}{l l|l l}
      \hline
      \# & Channel & \# & Channel \\
      \hline
      1 & $\pi^+ \pi^-\piz$  & 14 & $K^+ K^-\pi^+ \pi^-$ \\
      2 & $\pi^+ \pi^-2\piz$  & 15 & $K^+ K^-\pi^+ \pi^-\piz$ \\
      3 & $2\pi^+ 2\pi^-$  & 16 & $\Kpm\KS\pimp\pi^+\pi^-$ \\
      4 & $2\pi^+ 2\pi^-\piz$  & 17 & $K^+ K^-\eta$ \\
      5 & $\pi^+ \pi^-\eta$  & 18 & $K^+ K^-2\pi^+ 2\pi^-$ \\
      6 & $2\pi^+ 2\pi^-2\piz$  & 19 & $\Kpm\KS\pimp\pi^+\pi^-2\piz$ \\
      7 & $3\pi^+ 3\pi^-$  & 20 & $K^+ K^-2\pi^+ 2\pi^-\piz$ \\
      8 & $2\pi^+ 2\pi^-\eta$  & 21 & $K^+ K^-2\pi^+ 2\pi^-2\piz$ \\
      9 & $3\pi^+ 3\pi^-2\piz$  & 22 & $\Kpm\KS\pimp2\pi^+2\pi^-\piz$ \\
      10 & $4\pi^+ 4\pi^-$  & 23 & $K^+ K^-3\pi^+ 3\pi^-$ \\
      11 & $K^+ K^-\piz$  & 24 & $2K^+ 2K^-$ \\
      12 & $\Kpm\KS\pimp$  & 25 & $p \bar{p} \piz$ \\
      13 & $K^+ K^-2\piz$  & 26 & $p\bar{p} \pi^+ \pi^-$ \\
      \hline
    \end{tabular}
      \label{tab:hadronCombination}
        \end{center}
\end{table}

\begin{figure}
\includegraphics[width=\columnwidth]{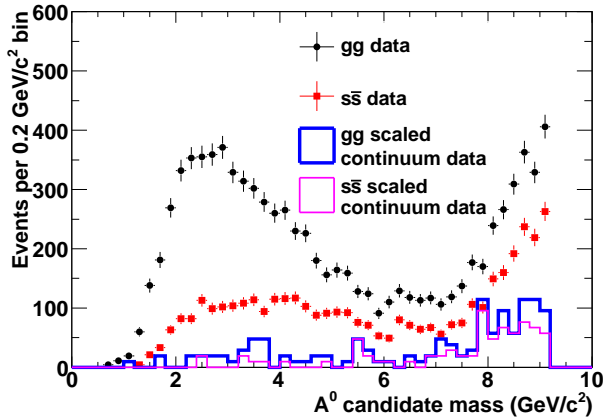}
\caption{
$A^0$ candidate mass spectra after applying all selection criteria. We reconstruct $A^0\to gg$ using the 26 channels listed in Table \ref{tab:hadronCombination} and $A^0\to \ssbar$ using the subset of the same 26 channels that includes two or four kaons. The $A^0$ candidate mass is the invariant mass of the reconstructed hadrons in each channel. The black points with error bars are on-resonance data for $A^0\to gg$. The red squares with error bars are on-resonance data for $A^0\to \ssbar$. The thick blue histogram is $A^0\to gg$ in off-resonance data normalized to the on-resonance integrated luminosity. The thin magenta histogram is $A^0\to \ssbar$ in off-resonance data normalized to the on-resonance integrated luminosity.
}
\label{fig:massSpec}
\end{figure}

We reconstruct $A^0\to gg$ using 26 channels as listed in Table \ref{tab:hadronCombination}. We do not use two-body decay channels because a $CP$-odd Higgs boson cannot decay into two pseudoscalar mesons. Charged kaons, pions, and protons are required to be positively identified. To reduce the number of misreconstructed candidates in an event, we require the number of reconstructed charged tracks in an event to match the number of charged tracks in the corresponding decay mode (including the \pip\pim). For example, we reconstruct ten-track events only as $K^+ K^-3\pi^+ 3\pi^-$, $\Kpm\KS\pimp2\pi^+2\pi^-\piz$ (two tracks from a \KS), or $4\pi^+ 4\pi^-$. The \piz and $\eta$ candidates are reconstructed from two photon candidates. The \KS candidates are reconstructed using two charged pions of opposite charge. We define our $A^0\to \ssbar$ sample as the subset of the 26 $A^0\to gg$ decay channels that include two or four kaons (channels 11--24 in Table \ref{tab:hadronCombination}). In simulated $A^0\to \ssbar$ events, there is a negligible contribution from channels that do not include at least two kaons. We form an $A^0$ candidate by adding the four-momenta of the hadrons. Similarly, we form an \OneS candidate by using the $A^0$ candidate and a photon with energy more than 200 MeV in the \epem CM frame. To improve the $A^0$ mass resolution, we constrain the photon and the $A^0$ candidates to have an invariant mass equal to the \OneS mass and a decay vertex at the beam spot. The $\chi^2$ probability of the constrained fit is required to be greater than $10^{-3}$. This rejects 77\% of the misreconstructed $A^0$ candidates, which includes candidates with misidentified charged kaons, pions, and protons. We reject \OneS candidates if the radiative photon, when combined with another photon in the event that is not used in the reconstruction of a \piz or $\eta$ candidate, has an invariant mass within 50\mevcc of the \piz mass. This removes backgrounds where a photon from a \piz decay is misidentified as the radiative photon. We also reject \OneS candidates if the Zernike moment $A_{42}$ \cite{ref:Zernike} of the radiative photon is greater than 0.1. This removes backgrounds where showers from both photons from a \piz decay overlap and are mistaken as the radiative photon. If there is more than one $\TwoS \to \pip\pim\OneS, \OneS \to \gamma A^0$ candidate that passes all the selection criteria in an event, the candidate with the highest product of MLP output and $\chi^2$ probability is kept. Figure \ref{fig:massSpec} shows the $A^{0}$ candidate invariant mass spectra for the $A^0\to gg$ and $A^0\to \ssbar$ channels separately after applying all selection criteria and selecting one candidate per event.

We use our off-resonance sample to estimate the continuum contribution in the on-resonance sample. Fifteen percent of the candidates in the on-resonance sample are determined to come from non-\TwoS decays.

We use simulated \TwoS events to study the remaining backgrounds, which originate mainly from $\OneS \to ggg$ and $\OneS \to \gamma gg$, where the gluons hadronize to more than one daughter. In $\OneS \to ggg$ decays, a \piz from the gluon hadronization is mistaken as the radiative photon. This decay mode contributes most of the background candidates with $A^0$ masses between 7 and 9\gevcc. The candidates with $A^0$ masses between 2 and 4\gevcc are mostly $\OneS \to \gamma gg$. CLEO measured the $\OneS\to \gamma f_2 (1270)$ \cite{ref:CLEO1270} and $\OneS\to \gamma f_2 '(1525)$ \cite{ref:CLEO1525} branching fractions. We do not expect these decays to be a background to the search for a narrow $A^{0}$ because they mainly decay to two-body final states and have decay widths of 100\mevcc.

To determine the number of signal events, we define a mass window, centered on the hypothesis $A^0$ mass, that contains 80\% of simulated signal events at that mass. For example, in simulated 3\gevcc $A^0 \to \ssbar$ events, 80\% of the events that pass the selection criteria have a reconstructed invariant mass for the $A^0$ within $\pm$170\mevcc of 3\gevcc. The mass windows are estimated for several $A^0$ masses for both $gg$ and \ssbar, and interpolated for all other masses. A sideband region is defined as half of the mass window size adjacent to both sides of the mass window. Again, for example, the lower sideband for a 3\gevcc $A^0 \to \ssbar$ would be from 2.66 to 2.83\gevcc, and the upper sideband would be from 3.17 to 3.34\gevcc.

Using simulated events, we estimate efficiencies of reconstructing the whole decay chain by taking the number of events in a signal mass window, subtracting the number of events in the sidebands, and dividing the difference by the number of simulated events. We interpolate the efficiencies for all hypothesis $A^0$ masses.

Our efficiency measurements of $gg$ and \ssbar into the 26 channels are dependent on the hadronization modelling by \mbox{\tt Jetset}. The accuracies of the simulated branching fractions of $gg$ and \ssbar to different final states are difficult to determine. We correct for this by comparing simulations with data in $\OneS \to \gamma gg$ decays. We count the number of events in the 26 channels where the reconstructed $gg$ mass is between 2 and 4\gevcc in data, and compare that to simulated $\TwoS \to \pip\pim\OneS, \OneS \to \gamma gg$ events in the same mass range. The background in this mass region is almost entirely from $\OneS \to \gamma gg$ decays. The number of $\OneS \to \gamma gg$ events is too few at masses above 4\gevcc to allow any meaningful study. For each of the 26 channels listed in Table \ref{tab:hadronCombination}, we calculate a weight that is the ratio of the event yields in data and simulation. We apply these weights to our efficiency calculations to determine how much the signal efficiency changes. The efficiencies change by a factor of 0.66 on average for $A^0 \to gg$ and 1.09 for $A^0\to\ssbar$. We correct the efficiencies by multiplying our measured efficiencies by these factors and assign an uncertainty due to hadronization modelling of ($1-0.66$)/0.66 = 50\% to all $A^0 \to gg$ and $A^0 \to \ssbar$ efficiencies since the correction is based on simulated $\OneS\to\gamma gg$ decays but not $\OneS\to\gamma \ssbar$ decays. We do not correct for, or assign hadronization modelling uncertainty to, $A^0 \to gg$ of invariant mass from 0.5 to 0.6 \gevcc because a $CP$-odd $A^0$ can decay to only \pip\pim\piz in that mass region.  Signal efficiencies range from 0.07 to $4\times 10^{-4}$ for $gg$ and 0.04 to $1\times 10^{-3}$ for \ssbar. The efficiencies are lower for higher $A^{0}$ masses because a more massive $A^{0}$ decays to more hadrons, which increases the probability of misreconstruction.

\begin{figure}
\includegraphics[width=\columnwidth]{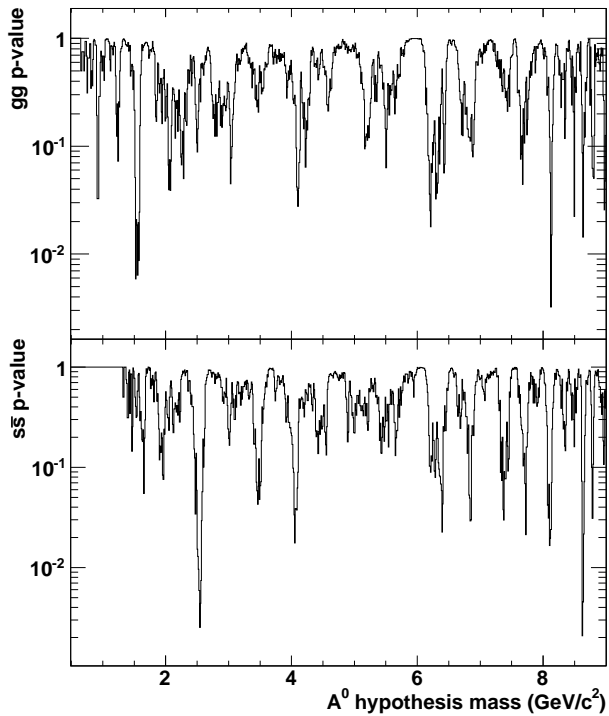}
\caption{
The probability of observing at least the number of signal events, assuming a null hypothesis for the existence of the decay $\OneS \to \gamma A^0 , A^0 \to gg$ (top), and $\OneS \to \gamma A^0 , A^0 \to \ssbar$ (bottom).
}
\label{fig:pValue}
\end{figure}

An $A^0$ signal would appear as a narrow peak in the candidate mass spectrum. To look for a signal, we scan the mass spectrum in 10\mevcc steps from 0.5\gev to 9.0\gevcc. Our null hypothesis is that the signal rate is 0 in the signal mass window. We use sidebands to estimate the number of background events in the signal region. Using Cousins' method \cite{ref:cousins}, we calculate a probability (p-value) of seeing the observed result or greater in the signal mass region given the null hypothesis. We do this separately for $A^0 \to gg$ and $A^0 \to \ssbar$. Figure \ref{fig:pValue} is the resulting p-value plot for all hypothesis masses. The minimum p-value for $A^0 \to gg$ is 0.003 and occurs at an $A^0$ mass of 8.13\gevcc. The minimum p-value for $A^0 \to \ssbar$ is 0.002 and occurs at an $A^0$ mass of 8.63\gevcc. These results are equivalent to Gaussian standard deviations of 2.7 and 2.9, respectively. We use $10^{4}$ simulated experiments to calculate how often such a statistical fluctuation might occur. For $A^0 \to gg$, 86\% of the simulated experiments have a minimum p-value less than 0.003. For $A^0 \to \ssbar$, 59\% of the simulated experiments have a minimum p-value less than 0.002. Therefore, we conclude that there is no evidence for the light $CP$-odd Higgs boson.

The dominant systematic uncertainty on the product branching fraction upper limit is related to the efficiency, which was described earlier in the text. Other systematic uncertainties, which are small compared to the 50\% uncertainty due to hadronization modelling, include Monte Carlo statistical uncertainties (1--7\%), efficiency variations in estimating the size of the mass windows (5\%), dipion branching fraction (2\%), \TwoS counting (1\%), and dipion selection efficiency (1\%). The systematic uncertainties are summed in quadrature and total 51\%.

\begin{figure}
\includegraphics[width=\columnwidth]{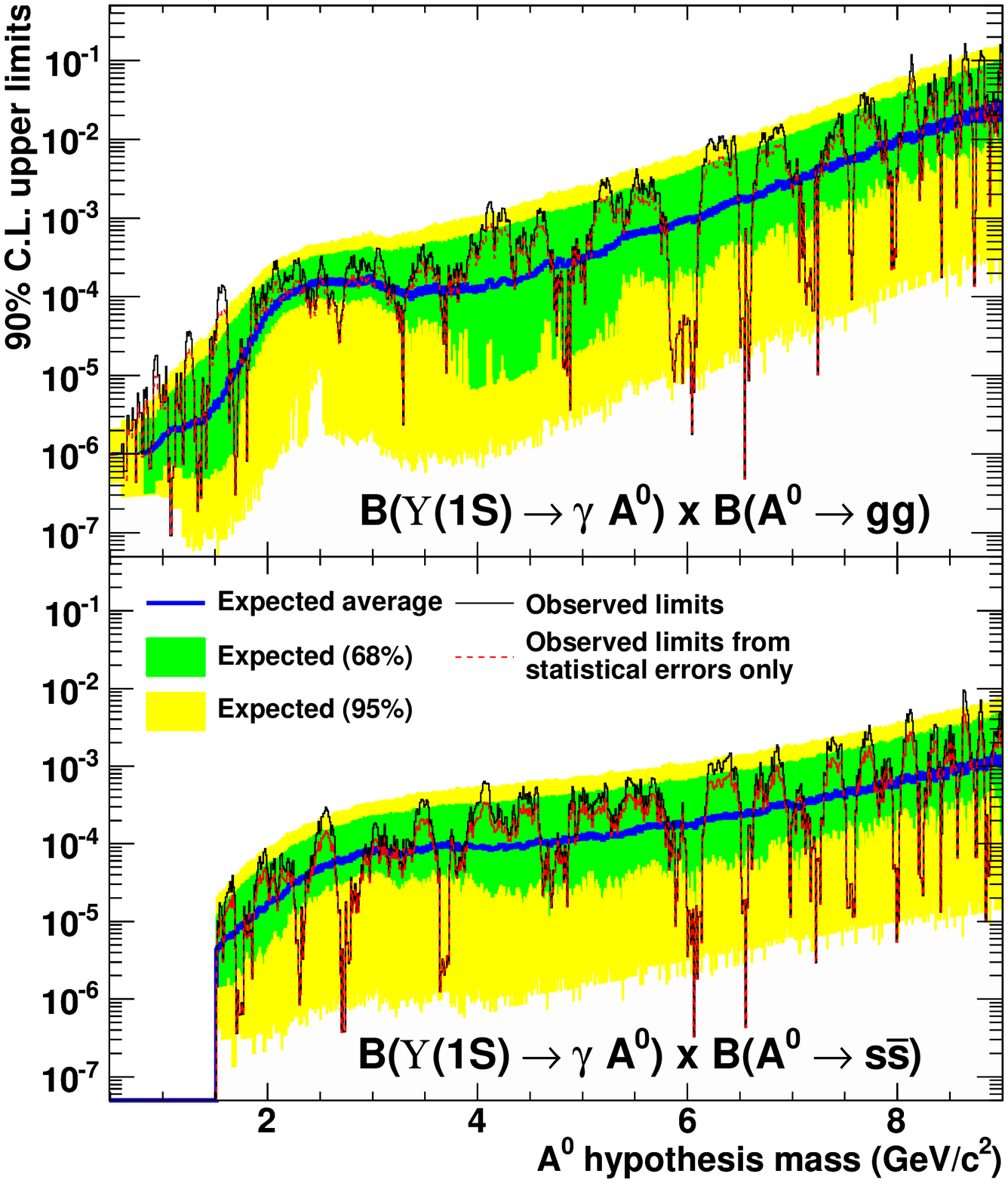}
\caption{
(color online) The 90\%-confidence-level upper limits (thin solid line) on the product branching fractions $\mathcal{B}(\OneS\to \gamma A^0) \cdot
\mathcal{B}(A^0 \to gg$) (top) and $\mathcal{B}(\OneS\to \gamma A^0) \cdot
\mathcal{B}(A^0 \to \ssbar$) (bottom). We overlay limits calculated using statistical uncertainties only (thin dashed line). The inner band is the expected region of upper limits in 68\% of simulated experiments. The inner band plus the outer band is the expected region of upper limits in 95\% of simulated experiments. The bands are calculated using all uncertainties. The thick line in the center of the inner band is the expected upper limits calculated using simulated experiments.
}
\label{fig:BFUL}
\end{figure}

We calculate 90\%-confidence-level (CL) upper limits (Fig. \ref{fig:BFUL}) on the product branching fractions $\mathcal{B}(\OneS\to \gamma A^0) \cdot
\mathcal{B}(A^0 \to gg$) and  $\mathcal{B}(\OneS\to \gamma A^0) \cdot
\mathcal{B}(A^0 \to \ssbar$) using a profile likelihood approach \cite{ref:Rolke}. We do this by calculating an upper limit of the mean number of signal events in the signal region given the number of events observed in the sidebands, and dividing by the efficiency, dipion branching fraction, and the number of \TwoS mesons produced. The number of background events is assumed to be Poissonian distributed and the efficiency distribution is assumed to be Gaussian with width equal to the total systematic uncertainty.

In summary, we select dipions in \TwoS decays to obtain a sample of \OneS mesons. We reconstruct the \OneS decay using a photon and a hadronic system. We observe no signals in the hadronic invariant mass spectra and set upper limits at 90\% CL on the product branching fractions $\mathcal{B}(\OneS\to \gamma A^0) \cdot
\mathcal{B}(A^0 \to gg$) from $10^{-6}$ to $10^{-2}$ and $\mathcal{B}(\OneS\to \gamma A^0) \cdot
\mathcal{B}(A^0 \to \ssbar$) from $10^{-5}$ to $10^{-3}$. We do not observe a NMSSM $A^0$ or any narrow hadronic resonance.

\input acknowledgements.tex

\end{document}

%% file: authors_jun2013.tex
%
\author{J.~P.~Lees}
\author{V.~Poireau}
\author{V.~Tisserand}
\affiliation{Laboratoire d'Annecy-le-Vieux de Physique des Particules (LAPP), Universit\'e de Savoie, CNRS/IN2P3,  F-74941 Annecy-Le-Vieux, France}
\author{E.~Grauges}
\affiliation{Universitat de Barcelona, Facultat de Fisica, Departament ECM, E-08028 Barcelona, Spain }
\author{A.~Palano$^{ab}$ }
\affiliation{INFN Sezione di Bari$^{a}$; Dipartimento di Fisica, Universit\`a di Bari$^{b}$, I-70126 Bari, Italy }
\author{G.~Eigen}
\author{B.~Stugu}
\affiliation{University of Bergen, Institute of Physics, N-5007 Bergen, Norway }
\author{D.~N.~Brown}
\author{L.~T.~Kerth}
\author{Yu.~G.~Kolomensky}
\author{M.~J.~Lee}
\author{G.~Lynch}
\affiliation{Lawrence Berkeley National Laboratory and University of California, Berkeley, California 94720, USA }
% ...
\collaboration{The \babar\ Collaboration}
\noaffiliation

%% file: abstract.tex
We search for the decay $\Y1S \to \gamma A^0 , A^0 \to gg$ or \ensuremath{s\overline s}\xspace, where $A^0$ is the pseudoscalar light Higgs boson predicted by the next-to-minimal supersymmetric standard model.
We use a sample of $(17.6 \pm 0.3)\times 10^6$  $\Y1S$ mesons produced in the $\mbox{\slshape B\kern-0.1em{\smaller A}\kern-0.1em
    B\kern-0.1em{\smaller A\kern-0.2em R}}$\  experiment via $\ensuremath{e^+e^-}\xspace \to \Y2S \to \ensuremath{\pi^+}\xspace\ensuremath{\pi^-}\xspace\Y1S$.
We see no significant signal and set 90\%-confidence-level upper limits 
on the product branching fraction 
$\mathcal{B}(\Y1S\to \gamma A^0) \cdot
\mathcal{B}(A^0 \to gg$ or $\ensuremath{s\overline s}\xspace$)
ranging from $10^{-6}$ to $10^{-2}$ for $A^0$ masses in the range 0.5 to 9.0\ensuremath{{\mathrm{\,Ge\kern -0.1em V\!/}c^2}}\xspace.  

%% file: acknowledgements.tex
We are grateful for the excellent luminosity and machine conditions
provided by our \pep2\ colleagues, 
and for the substantial dedicated effort from
the computing organizations that support \babar.
The collaborating institutions wish to thank 
SLAC for its support and kind hospitality. 
This work is supported by
DOE
and NSF (USA),
NSERC (Canada),
IHEP (China),
CEA and
CNRS-IN2P3
(France),
BMBF and DFG
(Germany),
INFN (Italy),
FOM (The Netherlands),
NFR (Norway),
MIST (Russia), and
PPARC (United Kingdom). 
Individuals have received support from CONACyT (Mexico), A.~P.~Sloan Foundation, 
Research Corporation,
and Alexander von Humboldt Foundation.